\documentstyle[preprint,aps]{revtex}
\tighten
\def\slash#1{#1\!\!\!/}    
\def\slashb#1{#1\!\!\!/}    

\begin{document}
\preprint{BIHEP-TH-97-10}
\title{Narrow width of a glueball decay into two mesons}
\author{
{Jun Cao,   Tao Huang   and    Hui-fang Wu}\\
{\small CCAST (World Laboratory) P.O. Box 8730, Beijing, 100080, P.R. China}\\
{\small Institute of High Energy Physics, P.O. Box 918(4), Beijing,
100039 P.R. China\thanks{Mailing address.} }
}
\date{\today}
\maketitle
\begin{abstract}
The widths of a glueball decay to two pions or kaons are analyzed in the pQCD
framework. Our results show that the glueball ground state has small
branching ratio for two-meson decay mode, which is around $10^{-2}$. The
predicted values are consistent with the data of $\xi\to\pi\pi, KK$ if $\xi$
particle is a $2^{++}$ glueball. Applicability of pQCD to the glueball decay
and comparison with $\chi_{cJ}$ decay are also discussed.
\end{abstract}
\section{Introduction}

\parindent 20pt
The existence of glueballs and hybrids is a direct
consequence of QCD. Since the $J/\psi$ is predicted to have appreciable
decays to $\gamma g g$ where two gluon formation is expected to enhance the
production of tensor and scalar glueballs, radiative $J/\psi$ and
$\psi^\prime$  decays have long been realized to be a favored area for
glueball searches. At present one pays particular attention to three states:
$f_0 (1500)$, $f_J (1710)$ ($J$=0 or 2), and $\xi(2230)$ $(J\geq 2)$. 
They are glueball candidates of $0^{++}$ or $2^{++}$ states.

\par
In particular, the BES collaboration discovered new, non-strange decay modes
of the $\xi(2230)$ state, such as $\xi\rightarrow\pi\pi$ and
$\xi\rightarrow{p}\overline{p}$ \cite{bes}. Compared with other mesons, the
$\xi(2230)$ has many distinctive properties\cite{hjzc}:
\par
(1) Flavor-symmetric decays to light hadrons. After removal of the phase
space factor, the probability for $\xi\rightarrow\pi^+\pi^-$ is of the same
order as that for $\xi\rightarrow{K^+K^-}$.
\par
(2) Copiously production in radiative $J/\psi$ decays. From the upper limit
of $1\times{10^{-4}}$ for 
$Br(\xi\rightarrow{p}\overline{p})Br(\xi\rightarrow K\overline{K})$ \cite{barnes,pdg}, 
where $K\overline{K}$ includes all kaon pairs, a lower bound
$3\times{10}^{-3}$ for the $Br(J/\psi\rightarrow\gamma\xi)$ can be estimated.
\par
(3) Narrow width. Both results from Mark III and BES show that the width of
the $\xi(2230)$ is only about 20 MeV. Assuming $\Gamma_\xi=20$ MeV, one can
easily estimate from (2) that the $Br(\xi\rightarrow{K^+K^-})$ and
$Br(\xi\rightarrow\pi^+\pi^-)$ are smaller than 2\%, resulting in partial
widths $\Gamma_{\pi^+\pi^-}$ and $\Gamma_{K^+K^-}$ smaller than 400 KeV 
\cite{hjzc}.

\par
As a consequence, the $q\overline{q}$ model, multi-quark model and hybrid model can
not easily reproduce three observations above. On the other hand, these properties
are naturally explained by identificating $\xi(2230)$ as a 
glueball state with $J^{PC}=2^{++}$.
\par
Since the observed decay modes into $\pi\pi$, $K\overline{K}$ and $p\overline{p}$ are
expected to be only a small portion of the decay modes of the $\xi(2230)$,
searches for other decay modes are very important. From theoretical point
of view, the narrow width of a glueball follows from a loose interpretation
of the OZI rule. The gluons in the glueball would annihilate and create a
$q\overline{q}$ pair that would form the lighter hadrons. Since this suppression
only acts at one vertex, it is so called $\sqrt{{\rm OZI}}$ rule 
\cite{robson}. For example, the total width of a $2^{++}$ glueball
can be estimated from $\sqrt{{\rm OZI}}$ rule to be about
$\sqrt{\Gamma_{f_2(1270)}\Gamma_{\chi_{c2}}}\simeq 20$ MeV.
\par
In order to understand the narrow decay width quantitatively, we study a
pure $0^{++}$ or $2^{++}$ glueball decay to two light mesons
in perturbative QCD (pQCD). Our results show that a pure glueball decay to
two light mesons has a small branching ratio or a narrow width, and this
conclusion can be generalized to be valid for any two mesons. As a
consequence, there is no dominant decay channel for a pure $0^{++}$ or
$2^{++}$ glueball.
\par
The paper is organized as follows. As a comparison with the glueball, the
formula for $\chi_{cJ}$ decay to two mesons in pQCD is reviewed briefly in
Sec. II. Glueball decays into two mesons are calculated in Sec. III. In Sec.
IV the numerical results and applicability of pQCD to glueball decays are
discussed. The last section is reserved for summary and conclusions.

\section{heavy quarkonium with $J^{PC}=0^{++}$ or $2^{++}$}
\par
It is interesting to compare a glueball decay with the $\chi_{cJ}$. A brief
review for $\chi_{cJ}$ decay will be given in the pQCD framework. 
For a heavy quarkonium coupling to gluons, the vertex can be obtained with
its radial wave function at origin, or its differentiation. Since the wave
function is sharply peaked at small internal momentum for heavy
quarkonium, the nonrelativistic limit is a good approximation. 
\par
As an example, we review the derivation of Br($\chi_{cJ}\rightarrow \pi\pi$)
in pQCD. To leading order, the Fermi movement of quarks in pion can be
neglected comparing with the momentum flow of order $m_c$, the mass of c
quark, in the hard scattering amplitude. Thus, the amplitude of
$\chi_{cJ}$ decay to two pions can be expressed in a factorizated form
\cite{cz} as $\pi$ form factor as
\begin{equation}\label{a}
A = \int d x d y \phi_\pi (x) T_H (x,y,Q^2) \phi_\pi (y) ,
\end{equation}
where $\phi_\pi (x)$ is the distribution amplitude of pion obtained by
integrating the transverse momentum of the Bethe-Salpeter wave function
\cite{2157}. The hard scattering amplitude $T_H (x,y,Q^2)$ (see fig.~1) will
include information of a heavy quarkonium coupling to two gluons:
\begin{equation}\label{th}
T_H = \left\{ \int \frac{d^4 k}{(2\pi)^4} Tr \left [\hat{O}_{\mu\nu}\chi (p,k)
\right] \right\} T^{\mu\nu} \frac{i g^4}{l_1^2 l_2^2} + 
\{x \leftrightarrow y \} ,
\end{equation} 
with $l_1$ and $l_2$ the momentum of gluons,
\begin{equation} \label{o}
\hat{O}^{\mu\nu} = \gamma^\mu \frac{1}{\slashb{p}+\slashb{k}-{\slash{l}}_1 -m}
\gamma^\nu + \{\mu \leftrightarrow \nu\} ,
\end{equation}
and the wave function of heavy quarkonium in nonrelativistic quark model
(NRQM) \cite{kuhn} 
\begin{equation} \label{chi}
\chi(p,k) = \sum_{M,S_z} (2\pi) \delta(k^0-\frac{\roarrow{k}^2}{2m})
\langle LM;SS_z|JJ_z\rangle \psi_{LM}(\roarrow{k}) P_{SS_z}(p,k) ,
\end{equation}
where $\psi_{LM} (\roarrow{k})$ and $P_{SS_z}(p,k)$ is spatial and spin part,
respectively.
$T^{\mu\nu}$ involve the spin part of pion wave function and present the
coupling of $g g \rightarrow \pi\pi$:
\begin{eqnarray} \label{tuv}
T^{\mu\nu}&=& Tr \left[ \gamma^\mu \frac{\gamma^5}{\sqrt{2}} \left (
\slashb{p}-\slashb{q} \right ) \gamma^\nu \frac{\gamma^5}{\sqrt{2}} \left (
\slashb{p}+\slashb{q} \right )\right ] \nonumber\\
&=& 4\left(p^\mu p^\nu -q^\mu q^\nu -g^{\mu\nu} m_c^2 \right ) .
\end{eqnarray}

For quarkonium in $P$-wave state, only $0^{++}$ and $2^{++}$ can decay to two
pions. In the case of $2^{++}$ state, the polarization tensor
$\epsilon^{\alpha\beta}$ is composed of the spin and orbital polarization
vectors, $\epsilon^\alpha(S_z)$ and $\epsilon^\beta(M)$, of $\chi_{c2}$ as
\begin{equation}
\epsilon^{\alpha\beta}(J_z)=\sum_{M,S_z}\langle 1M;1S_z|2 J_z\rangle 
\epsilon^{\alpha} (M)\epsilon^{\beta}(S_z) .
\end{equation}
The polarization tensor satisfies
$  \epsilon^{\alpha\beta}=\epsilon^{\beta\alpha} $,
 $ p^\alpha \epsilon^{\alpha\beta}=0 $,  $ {\epsilon^\alpha}_\alpha=0 $ 
and
\begin{equation} 
\sum_{J_z} \epsilon^{\alpha\beta}(J_z)\epsilon^{\mu\nu}(J_z) =
\frac{1}{2}({\cal P}^{\alpha\mu}{\cal P}^{\beta\nu}+{\cal P}^{\alpha\nu}
{\cal P}^{\beta\mu})-\frac{1}{3}{\cal P}^{\alpha\beta}{\cal P}^{\mu\nu},
\end{equation} 
where

\begin{equation} \label{pol}
{\cal P}^{\mu\nu}= - g^{\mu\nu}+\frac{p^\mu p^\nu}{m_c^2}.
\end{equation}
Now we arrive at the decay width if $\alpha_s(x_1x_2Q^2)\simeq\alpha_s(Q^2)$
\begin{equation} \label{c2part}
\Gamma(\chi_{c2}\rightarrow \pi\pi) =\frac{2 C^2}{5\pi^2 m_c^8}\left[
(4\pi\alpha_s)^2\phi_2^\prime(0) I_2 \right]^2,
\end{equation}
where $C^2=\frac{4}{27}$ is color factor and the integral $I_2$ is given by 
\begin{equation} 
I_2 =\int\frac{d x \phi_\pi(x)}{2 x(1-x)}\int\frac{d y \phi_\pi(y)}{2 y(1-y)}
\frac{1}{2(x+y-2 x y)}\left(1-\frac{(x-y)^2}{x+y-2 x y}\right) .
\end{equation} 
Similarly, The $0^{++}$ decay width can be expressed as
\begin{equation} \label{c0part}
\Gamma(\chi_{c0}\rightarrow \pi\pi) =\frac{4 C^2}{\pi^2 m_c^8}\left[
(4\pi\alpha_s)^2\phi_0^\prime(0) I_0 \right]^2,
\end{equation}
with the overlap integral $I_0$ given by
\begin{equation}
I_0 =\int\frac{d x \phi_\pi(x)}{2 x(1-x)}\int\frac{d y \phi_\pi(y)}{2 y(1-y)}
\frac{1}{2(x+y-2 x y)}\left(1+\frac{(x-y)^2}{2(x+y-2 x y)}\right).
\end{equation}
The total width of $\chi_{cJ}$ can be obtained by calculating the width of its
decay to two real gluons in a similar way. 
In NRQM, it is
\begin{mathletters}\label{ctot}
\begin{eqnarray} 
\Gamma_{\chi_{c2}} &=&\frac{8\alpha_s^2}{5m_c^4} |\phi_2^\prime(0)|^2 ,\\
\Gamma_{\chi_{c0}} &=&\frac{6\alpha_s^2}{m_c^4} |\phi_0^\prime(0)|^2 .
\end{eqnarray}
\end{mathletters}

Finally, we get the branching ratio
\begin{mathletters}
\begin{eqnarray}\label{cbr}
Br(\chi_{c2}\rightarrow\pi\pi) &=&\frac{16}{27}\frac{(4\pi\alpha_s
I_2)^2}{m^4} ,\\
Br(\chi_{c0}\rightarrow\pi\pi) &=&\frac{144}{81}\frac{(4\pi\alpha_s
I_0)^2}{m^4} ,
\end{eqnarray}
\end{mathletters}
which is independent of $\chi_{cJ}$ wave function.

\par
Numerical calculations can be easily done. However, the obtained results are
much smaller than the data \cite{h}. Even when stretching all parameters to
their extreme values, the predictions stay a factor 3-6 below the data
\cite{bk}. Moreover, the obtained branching ratio dominantly come from the
end-point region, where the pQCD is not available. As stated by Bodwin,
Braaten, and Lepage \cite{bbl}, the color-octet decay contribution arising
from the higher Fock component $|c\overline{c}g\rangle$ of the $\chi_{cJ}$
wave function is actually not suppressed with respect to the usual
color-singlet decay, owing to the angular momentum barrier in
$|c\overline{c}\rangle$ contribution, and is necessary to separate
rigorously short-distance effects and long-distance effects. Including the
color-octet contributions, branching ratio of $\chi_{cJ}\to\pi\pi$ can be
compared with the experimental data \cite{bk}.

\section{glueball}

\parindent 20pt
Similar to the $Q\overline{Q}$ state, we apply pQCD to the glueball decay. The
amplitude for its decay to two pions can be written as eq.~(\ref{a}), too.
Now $T_H$ contain the glueball wave function. The coupling of a
$J^{++}$ state to two gluons can be obtained from the requirements such as
linear in the polarization vectors of two gluons, $\epsilon_1$ and
$\epsilon_2$, and Lorentz and gauge invariant \cite{kuhn,cfl}. Keeping only
the leading twist term, the wave function can be written phenomenologically
as
\begin{eqnarray}
\Psi(2^{++}) &=& \frac{\delta^{ab}}{\sqrt{8}} \epsilon^{\mu\nu} G^1_{\mu\rho}
G^2_{\nu\rho} \phi_2(k) ,\\
\Psi(0^{++}) &=& \frac{\delta^{ab}}{\sqrt{8}} \frac{{\cal P}^{\mu\nu}}
{\sqrt{3}} G^1_{\mu\rho} G^2_{\nu\rho} \phi_0(k) ,
\end{eqnarray}
where $ \frac{\delta^{ab}}{\sqrt{8}} $  is color wave function,
 $ \epsilon^{\mu\nu} $  is the polarization tensor of $2^{++}$ glueball and 
 ${\cal P}^{\mu\nu}$ is defined in eq.~(\ref{pol}).
 $ G_{\mu\nu}^i =\epsilon_\mu^i k_\nu^i - k_\nu^i \epsilon_\mu^i, (i=1,2) $,
 $ \epsilon_\mu^i $  and
 $ k_\nu^i $
is the polarization vector and momentum of the $i$-th gluon, respectively.

\par
The leading order contribution to a $2^{++}$ glueball decay to two pions in
pQCD is shown in fig.~2. In center of mass frame, the amplitude can be
written as
\begin{equation} 
A = g_s^2 \int d z \phi_\pi(z) d z^\prime \phi_\pi(z^\prime) \phi_2 (k)
\frac{1}{k_1^2 k_2^2} \sum_{s_1 s_2} \epsilon_{\mu\nu}(s) G_{\mu\rho}^1
(s_1) G_{\nu\rho}^2(s_2) \epsilon^1_\sigma(s_1) \epsilon^2_\eta(s_2)
T^{\sigma\eta} ,
\end{equation}
where $T^{\sigma\eta}$ is the same as in eq.~(\ref{tuv}) and the color
factor will be included in the width formula. 
The momenta of gluons are fixed by that of quarks in pions, i.e.,
$k_1=z{Q_1}+z^\prime{Q_2}$ and $k_2=\overline{z}Q_1+\overline{z}^\prime{Q_2}$.
The width can be obtained
\begin{equation}\label{2part}
\Gamma(2^{++}\rightarrow \pi\pi) = C \frac{\pi}{30 M} \left|
\int d z d z^\prime
\phi_\pi(z) \phi_\pi(z^\prime) \phi_2(k_i)
\alpha_s \frac{z \overline{z} + z^\prime 
\overline{z}^\prime}{z\overline{z}z^\prime\overline{z}^\prime} \right|^2 ,
\end{equation}
where C is the color factor ${2}\over{9}$ and $\overline{z}=(1-z)$.
The total width of $2^{++}$, can be obtained by
its decay width to two on-shell gluons with $\roarrow{k}_i^2=M^2/4$:
\begin{equation}\label{2tot}
\Gamma_{2^{++}}
= \frac{M^3}{320\pi} |\phi_2(k_i)|^2 .
\end{equation}

\par
Similarly, we can obtain the width of $0^{++}$ decay to two pions
\begin{equation}\label{0part}
\Gamma(0^{++}\rightarrow \pi\pi) = C \frac{\pi}{12 M} \left|
\int d z d z^\prime
\phi_\pi(z) \phi_\pi(z^\prime) 
\phi_0(k_i)\alpha_s 
\frac{(z-z^\prime)^2+2z\overline{z}^\prime+2z^\prime\overline{z}
}{z\overline{z} z^\prime\overline{z}^\prime} \right|^2 , 
\end{equation}
and the total width
\begin{equation}\label{0tot}
\Gamma_{0^{++}} = \frac{3 M^3}{256\pi}|\phi_0(k_i)|^2 .
\end{equation}

\par
Comparing eqs.~(\ref{2part}) and (\ref{0part}) with eqs.
(\ref{c2part})-(\ref{ctot}), it can be shown that there are two different
ingredients between the glueball decay and the $P$-wave quarkonium decay. i)
There is no end-point singularity in the expression (\ref{2part}) and
(\ref{0part}) for the glueball decay since gluons in the glueball are
directly related to the glueball wave function; ii) Since we get the total
width by the decay to two on-shell gluons, it depends on the glueball wave
function at a particular momentum point. Especially, the point is located at
$\roarrow{k}^2 = M^2/4$. Intuitively, the wave function must be peaked at
low momentum, since a composite particle has little amplitude for existing
while its constituents are flying apart with large momentum. While the mass
$M$ is large, the obtained width decreases fast and depends on the shape of
the wave function drastically. On the other side, the next order
contributions to the total width, whose amplitudes are expected to have the
form $\alpha_s\int{d^4}k\phi(k)f(k)$ in general, will be lack of the
suppression of the wave function and not sensitive to the shape of the wave
function as long as $f(k)$ is smooth. It means that the higher order
contributions are important to the total width if the mass is large, since
the zero-th order contribution is suppressed strongly by wave function. In
another word, the pQCD evolution of the wave function to a particular scale
will be nontrivial.
\par
The formula for decay to a pair of kaons can be easily obtained by
substituting the distrubtion amplitude of pions by that of kaons. As a
result of isospin symmetry, the branching ratio to $\pi^0\pi^0$ is half of
the charged channel except that $\pi^0\pi^0$ in the final state can be formed
{\it via} QCD anomaly. This contribution is negligible due to the minor difference
between the mass of $u$-quark and $d$-quark. The anomaly
contributions to $\eta\eta$, $\eta\eta^\prime$ and
$\eta^\prime\eta^\prime$
channels are not necessarily small, since two soft gluons have large
possibility to form a $\eta$ ($\eta^\prime$) meson \cite{shifman}:
\begin{equation}
\langle0|\frac{3\alpha_s}{4\pi}G^a_{\mu\nu}{\tilde G}^{a\mu\nu}|\eta\rangle
=\sqrt{\frac{3}{2}}f_\pi m_\eta^2 .
\end{equation}
However, it is a low energy theorem. If the energy scale of the decay
process is high enough, the anomaly contribution is also negligible as it
emerges {\it via} QCD renormalization and is essentially a higher order
contribution in $\alpha_s$. For example, the branching ratio of
$\chi_{cJ}\to\eta\eta$ has no apparent enhancement comparing with
$\pi^0\pi^0$ channel. So the branching ratios of $\eta\eta$,
$\eta\eta^\prime$ and $\eta^\prime\eta^\prime$ channels are expected to have
the same order of $\pi\pi$ channel if pQCD dominate the decay process.

\section{Applicability of pQCD and Numerical results}
\indent
\par
Evaluation of eqs. (\ref{2part}$\sim$\ref{0tot}) will require the glueball
wave function, which we have little information about up to now. In the
following context we will use an oscillator wave function
\begin{equation}\label{wf}
\phi (k) = A_g \exp(-b_g^2 \roarrow{k}^2) ,
\end{equation}
with parameter appropriately chosen. If $0^{++}$ and $2^{++}$ glueballs have
similar wave functions in momentum space as argued in ref.~\cite{nsvz}, a
constraint on $b_g$ can be obtained from their total decay width. The $0^{++}$
glueball candidates $f_0(1500)$ and $f_0(1700)$ have widths $100\sim150$
MeV. The $2^{++}$ glueball candidate $\xi (2230)$ has width $\sim$ 20 MeV.
From our width formulae (\ref{2tot}) and (\ref{0tot}) we have
\begin{equation}
\frac{\Gamma_{\xi (2230)}}{\Gamma_{f_0 (1500)}} =
\frac{4M_\xi^3A_2^2}{15M_{f_0}^3A_0^2} 
\frac{\exp (-2 b_g^2 M_\xi^2)}{\exp (-2 b_g^2 M_{f_0}^2)} \simeq
\frac{20~{\rm MeV}}{120~{\rm MeV}}
\end{equation}
If the masses of scalar and tensor states are equal, the ratio will be
simply $\case{4}{15}$, which is in accord with ref.~\cite{cfl}. Taking into
account the mass gap, we get $b_g^2 \sim 0.4$~GeV$^{-2}$ if $A_2\simeq A_0$.
In general, $A_g$ is larger for a higher mass state. For example, $A_K$ is
larger that $A_\pi$ (see below). So we get $b_g^2 > 0.4$. According to the
discussion in the previous section, the higher order corrections may be
large and should be larger for a higher mass state. Thus we can say that the
lower limit of $b_g^2$ is 0.4 GeV$^{-2}$. Furthermore, the parameter $b_g$
is related to the radius of hadron. We know that $b^2\simeq 0.8$ GeV$^{-2}$
for a well established pion oscillator wave function
\cite{hms}. Assuming that all hadrons have similar size, $b^2$ will be in
the same range for glueball. The parameter $b^2$ for the glueball will be
taken in the range $0.6\leq b^2\leq 0.8$ in the following context.

\par
Before getting the numerical results it is necessary to discuss whether the
derived formula is applicable or reliable. The applicability of the pQCD
theory to exclusive processes at the present experimental energy region was
argued by Isgur and Smith \cite{is} by excluding the contributions of the
end-point regions where sub-leading (higher twist) terms are {\it a priori}
likely to be greater than the perturbative contribution. Recently, the
applicability of pQCD to the pionic form factor has been examined by cutting
end-point contribution~\cite{hs} and by including the effects of Sudakov
form factor~\cite{ls}. The first approach argues that the pQCD results are
self-consistent in the energy region where the contributions after cutting
dominate. The second one attempts to explain the suppression in the
end-point regions by including the effects of Sudakov form factor of the
quarks, which serves as a natural filter to pick out the hard contributions.
Two approaches give similar conclusion that pQCD is applicable to the pion
form factor as $Q^2>4$ GeV$^2$.  In fact, the applicability of pQCD to
exclusive processes differs from one process to another and it depends on
the end-point singularities. For example, the hard scattering amplitude of
the $\chi_{cJ}$ decay to two pions is more singular than the case of the
pion form factor. The obtained rate of $\chi_{cJ}\rightarrow\pi\pi$ depends
strongly on the end-point behavior of the pion wave function~\cite{end}. In
the case of the glueball decay, the hard scattering amplitude of glueball
decay to two pions is less singular than that of the pion form factor, and
it has a good behavior to ensure that the dominant contributions come from
the hard part. We can argue that pQCD is applicable for the glueball,
particularly, for the glueball candidate $\xi(2230)$. We will adopt the
first method which is simpler.

\par
In order to realize the condition
$\alpha_s(z\overline{z}M^2)<1$ in eqs.~(\ref{2part}) and
(\ref{0part}), we extend the parameterization of $\alpha_s(Q^2)$ by replacing
\cite{bhl}
\begin{equation}
\alpha_s(Q^2)=\frac{12\pi}{(33-2n_f) \ln(Q^2+M_0^2)/\Lambda^2} 
\end{equation}
to reflect the fact that at low $Q^2$ the transverse momentum intrinsic to
the bound state wave-function flow through all the propagators. The parameter
$M_0$ can be determined to ensure $\alpha_s(0)= 1$. The cut contribution,
obtained by cutting the integral limit to keep only hard contributions, say,
$\alpha_s < 0.7$, will be compared with the uncut contribution.

\par
In order to calculate the decay width we take the wave function of the pion
and the kaon as~\cite{bhl}
\begin{mathletters}
\begin{eqnarray}\label{wfpi}
\psi_\pi(x,k_\perp)&=&A_\pi\exp\left[
-b^2\frac{k_\perp^2+m_q^2}{x(1-x)}\right] ,\\
\psi_K(x,k_\perp)&=&A_K\exp\left[-b^2\left(\frac{k_\perp^2+m_q^2}{x}+
\frac{k_\perp^2+m_s^2}{1-x}\right)\right] ,
\end{eqnarray}
\end{mathletters}
where $m_q$ means the mass of $u$ or $d$ quark and $m_s$ the mass of $s$
quark. The distribution amplitude $\phi(x)$ can be obtained by integrating
$k_\perp$:
\begin{mathletters}
\begin{eqnarray}
\phi_\pi &=& \frac{A_\pi}{16\pi^2 b^2}x(1-x)
\exp\left[-b^2\frac{m_q^2}{x(1-x)}\right] ,\\
\phi_K &=& \frac{A_K}{16\pi^2 b^2} x(1-x)\exp\left[
-b^2\left( \frac{m_q^2}{x}+\frac{m_s^2}{1-x} \right) \right] .
\end{eqnarray}
\end{mathletters}
The parameters can be adjusted by using the constraints from decays
of $\pi\rightarrow\gamma\gamma$ and $\pi\rightarrow\mu\nu$
($K\rightarrow\mu\nu$) and the average quark transverse momentum
$\langle{k}_\perp^2\rangle_\pi\simeq\langle{k}_\perp^2\rangle_K\simeq$
(360 MeV)$^2$ \cite{hms,gh}. It turns out
$$
m_q =0.25~{\rm GeV},~~b^2 = 0.80~{\rm GeV}^{-2},~~A_\pi = 27.7~{\rm GeV}^{-1}
$$$$
m_s =0.55~{\rm GeV},~~b^2 = 0.80~{\rm GeV}^{-2},~~A_K =  51.0~{\rm GeV}^{-1} .
$$

\par
The numerical results are listed in the table I. The parameter $b_g$ is
taken as 0.6, 0.7 and 0.8. Here we haven't fixed the normalization constant
and the branching ratio is independent of $A_g^2$. The branching ratios of
$2^{++}$ and $0^{++}$ decay to $\pi^+\pi^-$ or $K^+K^-$ are around $10^{-2}$
for a pure glueball.  The numerical result is consistent with
$\xi\rightarrow\pi\pi$ if $\xi$ is a $2^{++}$ glueball.

\par
Also we examine the applicability of pQCD to the glueball decay by cutting
end-point contributions. Our numerical results show that the hard part
($\alpha_s<0.7$) contribute about 81\% to
$\Gamma(2^{++}\rightarrow\pi^+\pi^-)$, whose value is listed in table I, and
the ratio is nearly the same for different $b_g^2$ because the partial width
is not sensitive to $b_g^2$. For $0^{++}$ the ratio is about 50\%. Another
kind of distribution amplitude of pion, the asymptotic form, is also used.
We find that the branching ratio increases about 10\% for $2^{++}$ glueball
and 20\% for $0^{++}$ one and the hard contributions occupy about 61\% in
the width of $2^{++}$ decay to $\pi^+\pi^-$ and 30\% in the width of
$0^{++}$ decay.

\section{conclusions and summary}
\indent\par
We have analyzed the decay width of a glueball decay to two pions and kaons
in the pQCD framework and it can be generalized to other two-meson decay
channels. The numerical results of the decays show that the branching ratio
is small and the decay width is very narrow, compared to the $q\overline{q}$
bound state. The branching ratios are consistent with the data of
$\xi\rightarrow 2\pi, 2K$ if $\xi$ particle is a $2^{++}$ glueball.
\par
Our conclusions are as follows:
\par
(1) Any glueball candidate produced from $J/\psi$ radiative decay has small
branching ratio for two-meson decay mode, which is around
$10^{-2}$. This result is derived in the pQCD framework. However, if the mass of
glueball is not high enough to ensure that  pQCD dominate the decay process,
the QCD anomaly will play an important role and the $\eta\eta$,
$\eta\eta^\prime$, and $\eta^\prime\eta^\prime$ channels will enhance
apparently.
\par
(2) Applicability of pQCD to the glueball decay is discussed in our paper.
We show that the hard scattering amplitude of glueball decay has a good
behavior at the end-point region and is favored by pQCD while $\chi_{cJ}$
is opposite. The reason is that gluons are directly related to the wave
function for glueball decay. The
extra quark propagator (and the differentiation on it) produces the end-point
singularity in $\chi_{cJ}$ exclusive decay. More specific speaking, pQCD is
applicable for $\xi(2230)$ decay to two mesons if it is a $2^{++}$ glueball,
but is not reliable for $f_0(1500)$.
\par
(3) Related to the end-point singularity in $\chi_{cJ}$ decay, the color
octet contributions have the same order as color singlet and may be the
essential part for $P$-wave state quarkonium decay. Therefore, the dynamic
mechanism will be very different for decays of glueball and quarkonium with
the same $J^{PC}$.

\par
(4) It is not difficult to generalize our calculation to other two-meson
decay processes. The conclusion is expected to be similar. Therefore the
branching ratio is small for each decay mode and there is no dominant decay
channel for a pure glueball whose mass is high enough to ensure the pQCD
contributions dominate.

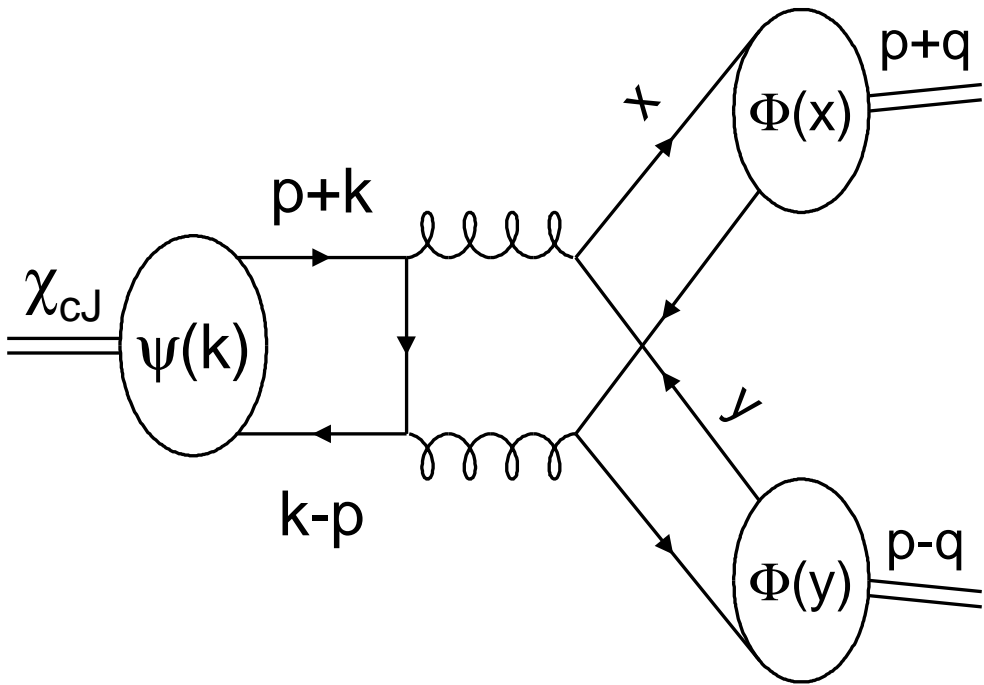
\begin{figure}
\caption{The hard scattering diagram for \protect$\chi_{cJ}\to\pi\pi$.
}
\label{fig1}
\end{figure}
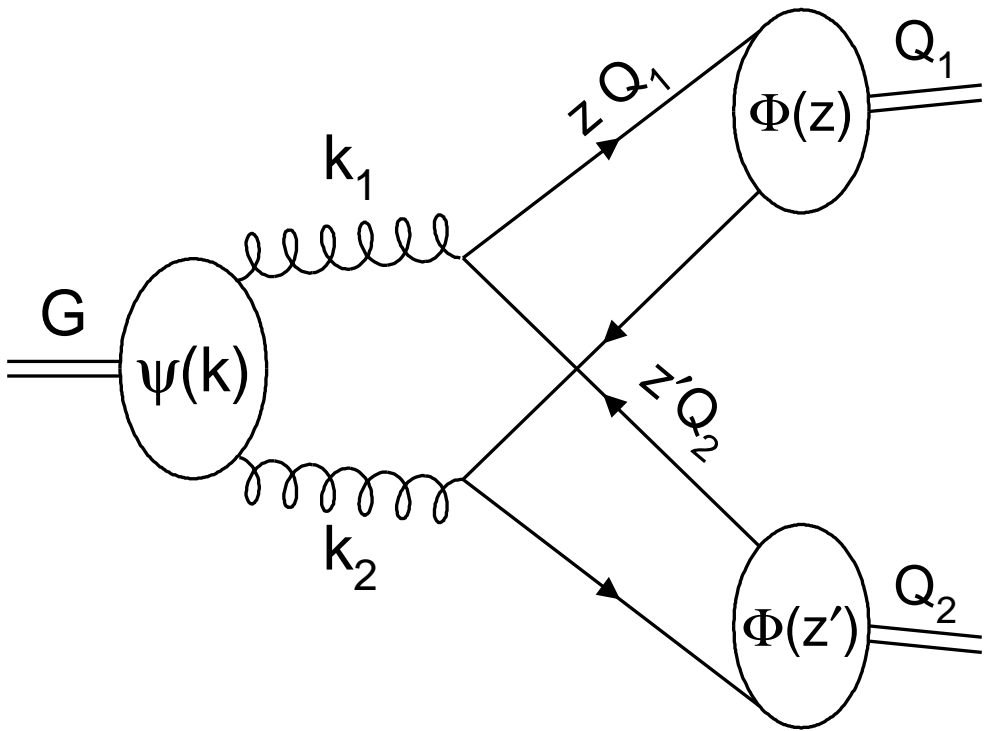
\begin{figure}
\caption{The hard scattering diagram for the glueball decay to  \protect$\pi\pi$.
}
\label{fig2}
\end{figure}
\begin{table}
\caption{Numerical results.}
\begin{tabular}{cccccc}
& $b_g^2$ & $\Gamma_{tot}$ & $\Gamma_{\pi^+\pi^-}$ & $Br(\pi^+\pi^-)$& $Br(K^+K^-)$ \\
\tableline
& 0.6 & 2.82$\times10^{-5}A_g^2$ & 1.08$\times10^{-7}A_g^2$ 
& 0.38$\times10^{-2}$ & 0.69$\times10^{-2}$\\
$\xi(2230)$ & 0.7 & 1.04$\times10^{-5}A_g^2$ & 1.01$\times10^{-7}A_g^2$ 
& 0.97$\times10^{-2}$ & 1.75$\times10^{-2}$\\
& 0.8 & 3.86$\times10^{-6}A_g^2$& 0.95$\times10^{-7}A_g^2$ 
& 2.45$\times10^{-2}$ & 4.45$\times10^{-2}$\\
\tableline
& 0.6 & 8.46$\times10^{-4}A_g^2$ & 6.00$\times10^{-6}A_g^2$ 
& 0.71$\times10^{-2}$ & 1.11$\times10^{-2}$\\
$f_0(1500)$ & 0.7 & 5.39$\times10^{-4}A_g^2$ & 5.64$\times10^{-6}A_g^2$ 
& 1.05$\times10^{-2}$ & 1.65$\times10^{-2}$\\
& 0.8 & 3.44$\times10^{-4}A_g^2$ & 5.32$\times10^{-6}A_g^2$ 
& 1.55$\times10^{-2}$ & 2.47$\times10^{-2}$
\end{tabular}
\end{table}

\end{document}